\documentclass[draft]{agujournal2019}
\usepackage{url} 
\usepackage[inline]{trackchanges} 
\usepackage{soul}

\usepackage{tabularx}
\usepackage{booktabs}
\usepackage{amsmath}
\usepackage{multirow}
\usepackage{bm}
\usepackage{subfigure}
\usepackage{amssymb}
\usepackage{chngpage}
\usepackage{array}
\usepackage{xcolor}
\draftfalse

\journalname{Radio Science}

\begin{document}

%
%


\title{Satellite-Terrestrial Channel Characterization in High-Speed Railway Environment at 22.6 GHz}

%
%




\authors{Lei Ma$^1$$^2$, Ke Guan$^1$$^2$, Dong Yan$^1$$^2$, Danping He$^1$$^2$, Nuno R. Leonor$^3$, Bo Ai$^1$$^2$, Junhyeong Kim$^4$$^5$}


\affiliation{1}{State Key Laboratory of Rail Traffic Control and Safety, Beijing Jiaotong University, Beijing, China}
\affiliation{2}{Beijing Engineering Research Center of High-Speed Railway Broadband Mobile Communications, China}
\affiliation{3}{Instituto de Telecomunica\c{c}\~{o}es, Leiria, Portugal and Polytechnic Institute of Leiria, Leiria, Portugal}
\affiliation{4}{Moving Wireless Network Research Section, Electronics and Telecommunications Research Institute (ETRI), Daejeon 34129, South Korea}
\affiliation{5}{School of Electrical Engineering, Korea Advanced Institute of Science and Technology (KAIST), Daejeon 34141, South Korea}




\correspondingauthor{Ke Guan}{kguan@bjtu.edu.cn}




\begin{keypoints}
\item Channel characterization
\item Millimeter wave channel
\item Ray tracing simulation
\item Railway communication
\item Satellite-terrestrial communication
\end{keypoints}

%
%

%
%


\begin{abstract}
The integration of satellite and terrestrial communication systems plays a vital role in the fifth-generation mobile communication system (5G) for the ubiquitous coverage, reliable service and flexible networking.
Moreover, the millimeter wave (mmWave) communication with large bandwidth is a key enabler for 5G intelligent rail transportation.
In this paper, the satellite-terrestrial channel at 22.6 GHz is characterized for a typical high-speed railway (HSR) environment.
The three-dimensional model of the railway scenario is reconstructed and imported into the Cloud Ray-Tracing (CloudRT) simulation platform.
Based on extensive ray-tracing simulations, the channel for the terrestrial HSR system and the satellite-terrestrial system with two weather conditions are characterized, and the interference between them are evaluated.
The results of this paper can help for the design and evaluation for the satellite-terrestrial communication system enabling future intelligent rail transportation.
\end{abstract}

%
%

%


%
%
%
%

\section{Introduction}
Satellite communication systems have been greatly developing in the domain of broadcasting, navigation, rescue, and disaster relief because of their potentiality to provide wide coverage and achieve high data rate transmission \cite{7230282}.
For most cases in previous generations, satellite systems were considered completely independent from the terrestrial communication \cite{8795462}. However, there exists a potential shortcoming that the satellite system will degenerate in the presence of shadowing, which occurs when the line-of-sight (LOS) link between the satellite and the terrestrial user is blocked by obstacles \cite{8081808,caixuesong2}.

The high-speed railway (HSR) is one of the most challenging scenarios in the fifth-generation mobile communication system (5G) \cite{Liuyu1}, whose demands for high data rate transmission and high reliability services have grown rapidly \cite{Liuyu2}.
On the one hand, the terrestrial cellular network can provide low cost coverage for high reliability applications in HSR environment through its non-LOS (NLOS) communication.
On the other hand, the satellite system can mitigate the problems of overload and congestion by providing wide coverage to complement and extend the dense terrestrial cells, especially for the terrestrial wireless communication in HSR areas.
Thus, it can be seen that a/an hybrid/integrated satellite-terrestrial cooperative communication system can realize genuine ubiquitous coverage for future HSR communications \cite{2011MIMO,7105655}.
In this case, the terrestrial mobile users can make full use of the spatial diversity gain by receiving independent multipath fading signals from satellites and terrestrial base stations. Consequently, the effectiveness and reliability of transmission will be greatly increased \cite{2011111}.

In order to realize this vision, the integrated satellite-terrestrial communication systems should be carefully designed, especially by analyzing the interference between the satellite link and the terrestrial link based on realistic channel characteristics \cite{fanwei2,fanwei3}.
Hence, it is essential to capture joint channel characteristics in cooperative satellite-terrestrial systems in order to make realistic performance assessments and channel modeling for future intelligent rail transportation \cite{ZhouTao1,ZhouTao3}.

Most of the existing performance analysis for satellite and terrestrial communication systems were studied based on pure mathematical models \cite{7343438,7373246,7308010,7156170,ZTE}.
However, the existing results can hardly be applied for the satellite-terrestrial channel characterization for HSR due to the particular geometrical and physical characteristics of the HSR environment.
Therefore, there is a lack of deep investigation for satellite-terrestrial channel based on a typical HSR scenario, which causes limited accuracy on coverage prediction and interference analysis.
Thus, in this study, we characterize the satellite-terrestrial channel at 22.6 GHz band comprehensively through simulation and modeling, with the following contributions:
\begin{itemize}
\item We reconstruct a typical HSR model and conduct extensive ray-tracing (RT) simulations in four terrestrial and satellite-terrestrial communication links with two weather conditions. Based on RT simulation results, the four links are characterized in terms of key channel parameters, containing root-mean square (RMS) delay spread (DS), Rician $K$-factor (KF), azimuth angular spread of arrival/departure (ASA/ASD), elevation angular spread of arrival/departure (ESA/ESD).
\item We predict the propagation behavior of all the objects in the simulation scenario, as well as the interaction between them. Besides, through calculating the signal-to-interference ratio (SIR), the interference between terrestrial HSR system and satellite-terrestrial system is evaluated.
\end{itemize}

The rest of this paper is organized as follows: Section II addresses the scenario modeling and the simulation setup. Key channel parameters are analyzed and characterized in Section III. Finally, conclusions and further work are drawn in Section IV.

\section{Railway Environment Reconstruction and Simulation Setup}
\subsection{Antenna Model}
\begin{figure}[!t]
\center
\includegraphics[width=1\columnwidth,draft=false]{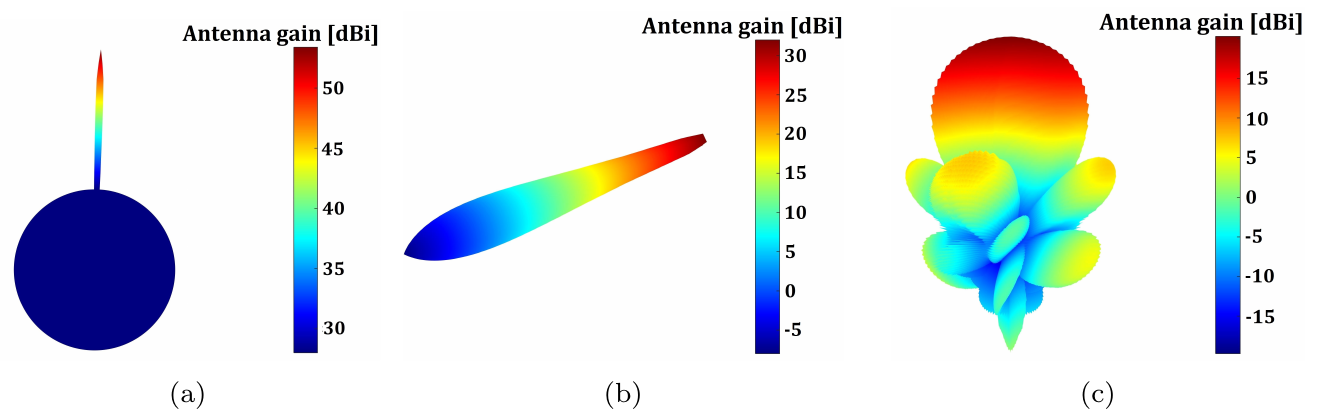}\\
\caption{Antenna patterns used in ray-tracing simulations: a) Satellite antenna pattern; b) Satellite UE antenna pattern; and c) Terrestrial antenna pattern.}
\label{fig:Antenna}
\end{figure}

For the satellite-terrestrial system, the satellite transmitter (Tx) employs the antenna called APSMLA609V01 which is provided by ITU, and the receiver (Rx) antenna is selected according to ITU-R S.456-6 \cite{465-6}. The antenna patterns of the satellite-terrestrial system are depicted in Fig. \ref{fig:Antenna}(a) and Fig. \ref{fig:Antenna}(b).
As for the terrestrial HSR system, both the Tx and Rx employ the same antenna. The antenna pattern is depicted in Fig. \ref{fig:Antenna}(c).

\subsection{Simulation Scenario}
\begin{figure}[!t]
	\center
	\includegraphics[width=0.7\columnwidth,draft=false]{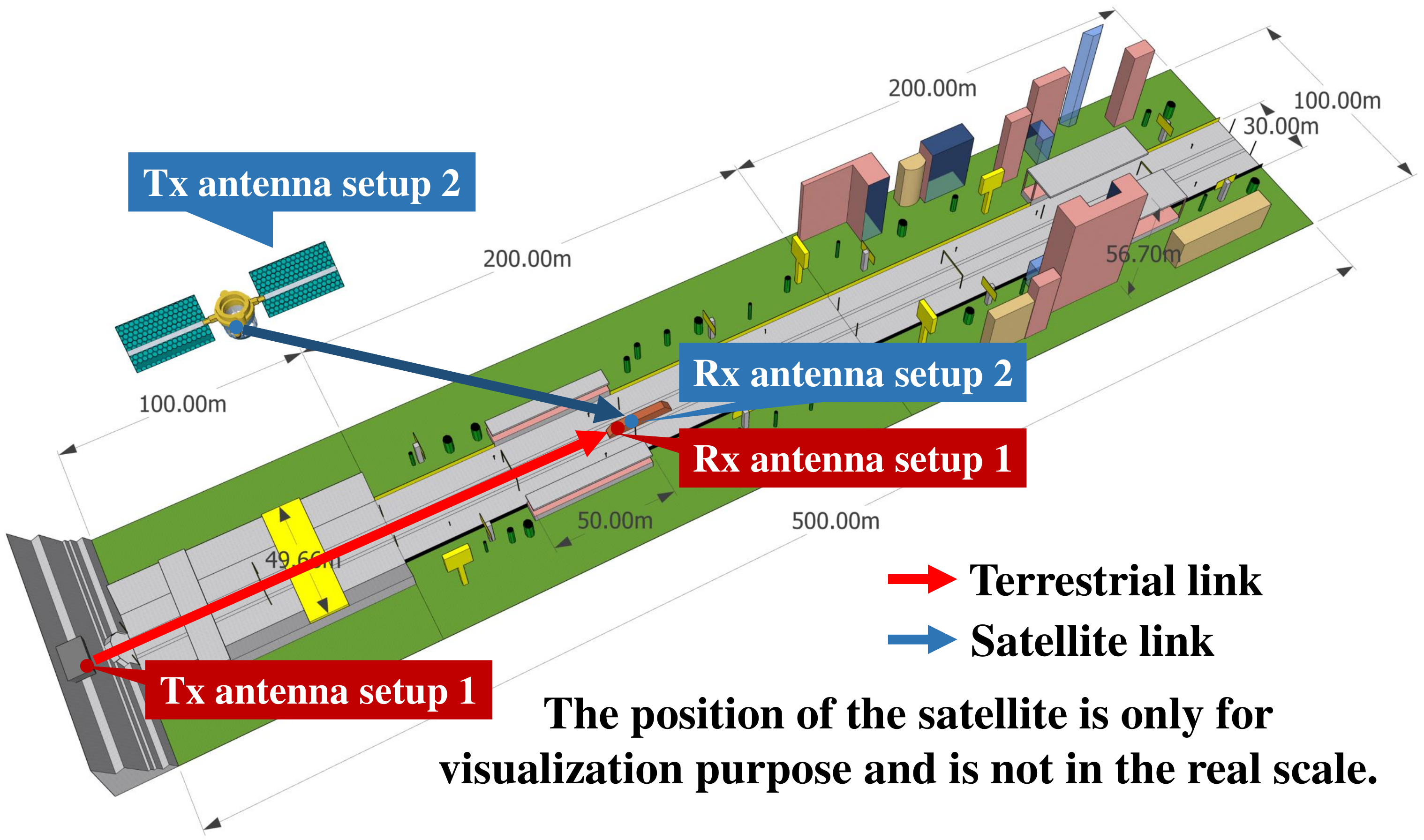}\\
	\caption{3D model of the railway scenario for ray-tracing simulation}
	\label{fig:3D model}
\end{figure}

For characterizing the channel in a HSR environment, it is important to define the scenario model with distinguished propagation features. In this paper, the three-dimensional (3D) model of a typical HSR scenario reconstructed via SketchUp tool is depicted in Fig. \ref{fig:3D model} \cite{8319730}.
Several buildings, traffic signs, billboards, crossing bridges, train stations, etc., are defined in the vicinity of the railway, composing a realistic HSR scenario.
All the objects are modeled according to their typical geometric shapes, and different materials are assigned to their surfaces.
Therefore, the electromagnetic behavior of each object could be effectively established, and the realistic description of the HSR environment allows the analysis of key channel parameters.

The train travels at a constant speed of 300 km/h during the 500 m movement in the scenario. 1441 samples are extracted in the simulation, corresponding to a sampling distance of 0.347 m.
For satellite-terrestrial links, the Tx is located on a geosynchronous satellite (GEO) called Koreasat 6, which is positioned at 116$^{\circ}$E overhead the equator and at a distance of approximately 37470 km from the target HSR scenario. The Rx is assembled to the rear of the train with a total height of 5.2 m, which includes the train height (4.5 m) and the antenna bracket.

\begin{figure}[!t]
\center
\includegraphics[width=0.65\columnwidth,draft=false]{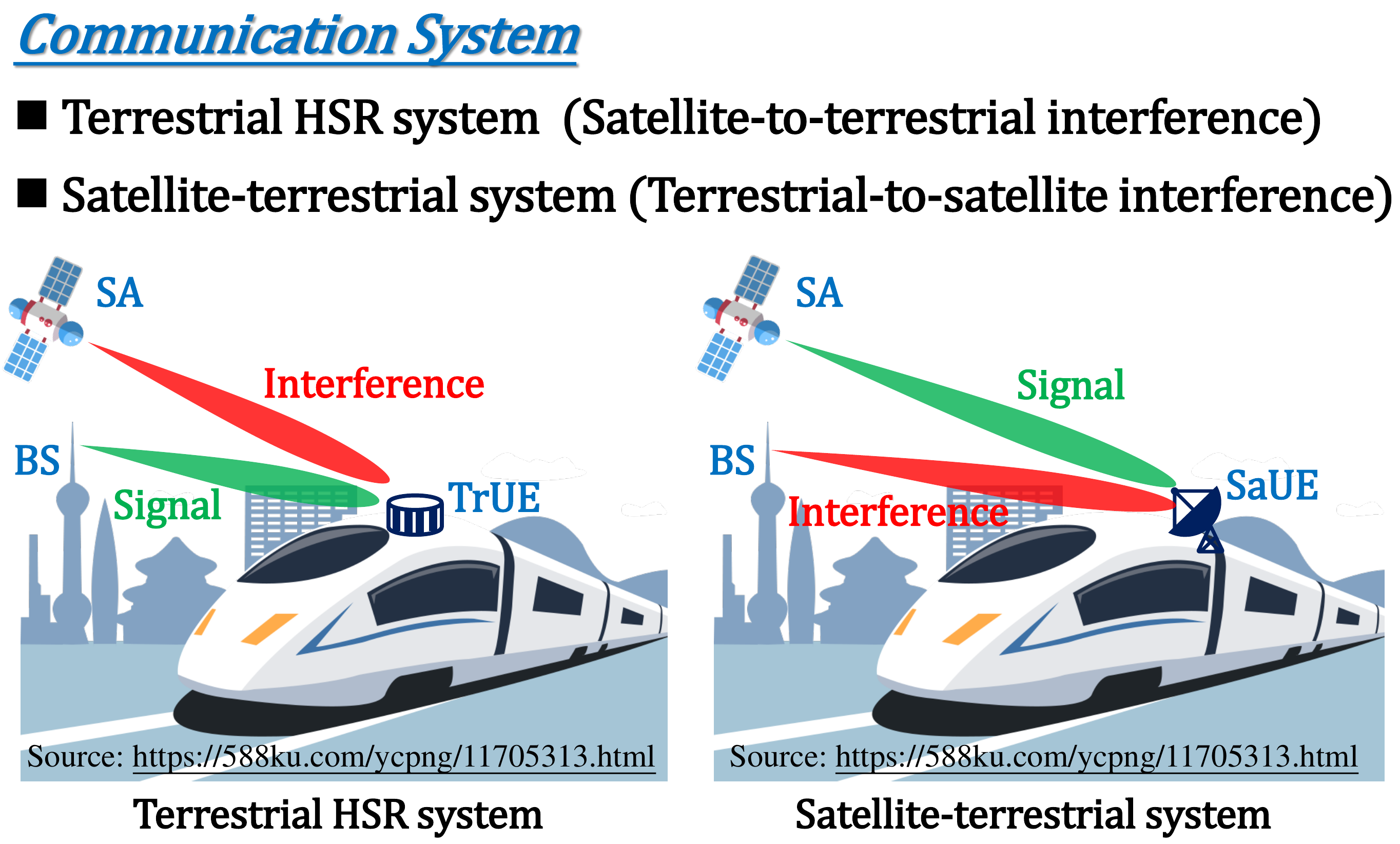}\\
\caption{Communication links for interference analysis}
\label{fig:Communication links}
\end{figure}

The presence of terrestrial links will affect the SIR of satellite-terrestrial links, and vice versa. Hence, for the SIR analysis between the terrestrial HSR system and satellite-terrestrial system, an additional communication link is included in the simulation. The terrestrial Tx is placed at the top of the steep wall with a height of 26 m, and the Rx is similarly assembled to the rear of the train with a total height of 4.7 m. Both the terrestrial and satellite-terrestrial links are depicted in Fig. \ref{fig:Communication links}.
The abbreviations in this figure are noted as follows: BS and TrUE are short for base station (i.e. Tx) and train user equipment (i.e. Rx) for the terrestrial HSR system, respectively. SA and SaUE are short for satellite antenna (i.e. Tx) and satellite user equipment (i.e. Rx) for the satellite-terrestrial system, respectively.

Table \ref{Table:scenario configuration} conclude the scenario configurations for the terrestrial HSR system and the satellite-terrestrial system. The communication scenarios for both systems are the same. The most differences are the locations of Tx and the selected antennas for Tx and Rx.

\begin{table}[!t]
\centering
\caption{Scenario configurations for terrestrial HSR system and satellite-terrestrial system}
\label{Table:scenario configuration}
\begin{tabular}{c|c|l|l}
\specialrule{0.3pt}{2pt}{0.5pt}
\specialrule{0.3pt}{0.5pt}{2pt}
Frequency                    & \multicolumn{3}{l}{22.1-23.1 GHz}                                \\ \hline\rule{0pt}{8pt}
Bandwidth                    & \multicolumn{3}{l}{1 GHz}                                        \\ \hline\rule{0pt}{8pt}
Antenna                     & \multicolumn{3}{l}{Directional antenna}                          \\ \hline\rule{0pt}{8pt}
\multirow{8}{*} {Terrestrial HSR system} & \multirow{4}{*}{Tx} & Power                               & 20 dBm                     \\ \cline{3-4}\rule{0pt}{8pt}
                             &                     & Maximum antenna gain                 & 16 dBi                     \\ \cline{3-4}\rule{0pt}{8pt}
                             &                     & Antenna beamwidth                    & 20 degree                  \\ \cline{3-4}\rule{0pt}{8pt}
                             &                     & Height                               & 26 m                       \\ \cline{2-4}\rule{0pt}{8pt}
                             & \multirow{3}{*}{Rx} & Maximum antenna gain                 & 22 dBi                     \\ \cline{3-4}\rule{0pt}{8pt}
                             &                     & Antenna beamwidth                    & 20 degree                  \\ \cline{3-4}\rule{0pt}{8pt}
                             &                     & Height                               & 4.7 m                      \\ \hline\rule{0pt}{8pt}
\multirow{8}{*}{Satellite-terrestrial system}   & \multirow{4}{*}{Tx} & Power                                & 40.6 dBm                     \\ \cline{3-4}\rule{0pt}{8pt}
                             &                     & Maximum antenna gain                 & 53 dBi                     \\ \cline{3-4}\rule{0pt}{8pt}
                             &                     & Antenna beamwidth                    & 1 degree                  \\ \cline{3-4}\rule{0pt}{8pt}
                             &                     & Height                               & 37469.3 km                       \\ \cline{2-4}\rule{0pt}{8pt}
                             & \multirow{3}{*}{Rx} & Maximum antenna gain                 & 32 dBi                     \\ \cline{3-4}\rule{0pt}{8pt}
                             &                     & Antenna beamwidth                    & 3 degree                  \\ \cline{3-4}\rule{0pt}{8pt}
                             &                     & Height                               & 5.2 m                      \\
\specialrule{0.3pt}{1pt}{0.5pt}
\specialrule{0.3pt}{0.5pt}{2pt}
\end{tabular}
\end{table}

Furthermore, the rainfall can significantly affect the performance of the wireless communication system since it causes additional attenuation to wave propagation, especially for the satellite-terrestrial link \cite{618-13}. Therefore, both communication systems are characterized for rainy and sunny weather conditions. In total, the simulation contains 8 cases which are summarized in Table \ref{Table:Analysis cases}. Suffix ``-R'' and ``-S'' represent the rainy and sunny weather condition, respectively.

\begin{table}[!t]
\centering
\caption{Analysis cases for the satellite-terrestrial channel}
\label{Table:Analysis cases}
\small
\begin{tabular}{c|c|c|c|c|c}
\specialrule{0.3pt}{2pt}{0.5pt}
\specialrule{0.3pt}{0.5pt}{2pt}
\textbf{Tx}         & \textbf{Rx}           & \textbf{Weather} & \textbf{Signal} & \textbf{Interference} & \textbf{Terminology} \\ \hline\rule{0pt}{8pt}
		\multirow{4}{*}{BS} & \multirow{2}{*}{TrUE} & Rainy            & \checkmark               &                       & BS2TrUE-R    \\ \cline{3-6}\rule{0pt}{8pt}
		&                       & Sunny            & \checkmark               &                       & BS2TrUE-S    \\ \cline{2-6}\rule{0pt}{8pt}
		& \multirow{2}{*}{SaUE} & Rainy            &                 & \checkmark                     & BS2SaUE-R    \\ \cline{3-6}\rule{0pt}{8pt}
		&                       & Sunny            &                 & \checkmark                     & BS2SaUE-S    \\ \hline\rule{0pt}{8pt}
		\multirow{4}{*}{SA} & \multirow{2}{*}{SaUE} & Rainy            & \checkmark               &                       & SA2SaUE-R    \\ \cline{3-6}\rule{0pt}{8pt}
		&                       & Sunny            & \checkmark               &                       & SA2SaUE-S    \\ \cline{2-6}\rule{0pt}{8pt}
		& \multirow{2}{*}{TrUE} & Rainy            &                 & \checkmark                     & SA2TrUE-R    \\ \cline{3-6}\rule{0pt}{8pt}
		&                       & Sunny            &                 & \checkmark                     & SA2TrUE-S    \\
		\specialrule{0.3pt}{1pt}{0.5pt}
		\specialrule{0.3pt}{0.5pt}{2pt}
	\end{tabular}
\end{table}

\subsection{Simulation Setup}
As a deterministic modeling method, RT simulations can provide full information of multipath effects in multiple domains and build accurate site-specific channel models.
It has been successfully used for different applications \cite{ZhouTao2,fanwei1,Guanke2,Guanke3,caixuesong1,caixuesong3,chenxiaoming}.
The RT simulator employed in this study, CloudRT, is jointly developed by Beijing Jiaotong University and Technische Universit{\"a}t Braunschweig.
It can trace rays corresponding to various propagation mechanisms, such as direct rays, reflected rays, scattered rays, etc., and is validated and calibrated by a large number of measurements at sub-6 GHz \cite{Abbas2015Simulation} and terahertz (THz) band \cite{Priebe2013Stochastic}.
More than ten properties of each ray can be output from RT results, such as reflection order, time of arrival, received power, AoA, AoD, EoA, EoD, etc.
More information on the CloudRT can be found in tutorial \cite{hedanping} and at http://www.raytracer.cloud/.
The setup for RT simulations is detailed in Table \ref{Table:SimulationSetUp}.
\begin{table}[!t]
\centering
\caption{Ray-tracing simulation setup}
\label{Table:SimulationSetUp}
\small
\begin{tabular}{c|l|l|l}
\specialrule{0.3pt}{2pt}{0.5pt}
\specialrule{0.3pt}{0.5pt}{2pt}
\multirow{5}{*}{\begin{tabular}[c]{@{}c@{}}Propagation\\ mechanism\end{tabular}}
& Direct                    & \multicolumn{2}{l}{\checkmark}\\ \cline{2-4}\rule{0pt}{8pt}
& Reflection                & \multicolumn{2}{l}{up to the 2$^{nd}$ order}\\ \cline{2-4}\rule{0pt}{8pt}
& Diffraction               & \multicolumn{2}{l}{Uniform theory of diffraction (UTD)}      \\ \cline{2-4}\rule{0pt}{8pt}
& Scattering                & \multicolumn{2}{l}{Directive scattering model} \\ \cline{2-4}\rule{0pt}{8pt}
& Transmission              & \multicolumn{2}{l}{\checkmark}      \\ \hline\rule{0pt}{8pt}
\multirow{5}{*}{Material}    & Building         & \multicolumn{2}{l}{Marble, Toughened glass} \\ \cline{2-4}\rule{0pt}{8pt}
& Steep wall, Cutting walls & \multicolumn{2}{l}{Brick} \\ \cline{2-4}\rule{0pt}{8pt}
& Railway furniture, Train  & \multicolumn{2}{l}{Metal}              \\ \cline{2-4}\rule{0pt}{8pt}
& Tree                      & \multicolumn{2}{l}{Wood}               \\ \cline{2-4}\rule{0pt}{8pt}
& Ground                    & \multicolumn{2}{l}{Concrete}           \\
\specialrule{0.3pt}{1pt}{0.5pt}
\specialrule{0.3pt}{0.5pt}{2pt}
\end{tabular}
\end{table}

Table \ref{Table:EM parameter} summarizes electromagnetism (EM) parameters of the involved materials, where $\varepsilon _{r}^{'}$ is the real part of the relative permittivity, $tan\delta $ is the loss tangent, $S$ and $\alpha$ are the scattering coefficient and scattering exponent of the directive scattering model \cite{Vittorio}\cite{wanglonghe2019hindawi}.
Particularly, parameters of the wood and concrete are calibrated \cite{Wang}.
\begin{table}[!t]
\centering
\caption{EM parameters of different materials}
\label{Table:EM parameter}
\small
\begin{tabular}{c|c|c|c|c|c|c}
\specialrule{0.3pt}{2pt}{0.5pt}
\specialrule{0.3pt}{0.5pt}{2pt}
Material & Marble & Toughened glass & Brick & Metal & Wood & Concrete \\ \hline
$\varepsilon _{r}^{'}$    & 3.0045     & 1.0538     & 1.9155    & 1             & 6.6           & 5.4745            \\ \hline
$tan\delta$               & 0.2828     & 23.9211    & 0.0568    & 10$^{7}$      & 0.9394        & 0.0021            \\ \hline
$S$                       & 0.0022     & 0.0025     & 0.0019    & 0.0026        & 0.0086        & 0.0011            \\ \hline
$\alpha$                  & 15.3747    & 5.5106     & 49.5724   & 17.7691       & 13.1404       & 109               \\
\specialrule{0.3pt}{1pt}{0.5pt}
\specialrule{0.3pt}{0.5pt}{2pt}
\end{tabular}
\end{table}

\section{Excess Propagation Attenuation}
Apart from the attenuation due to classic propagation mechanisms already in the CloudRT, the additional propagation attenuation caused by several effects is significant for millimeter wave (mmWave) communication links, which must be considered accordingly and added into the CloudRT in this study.
\subsection{Excess Propagation Attenuation for Terrestrial Links}
For terrestrial links, the attenuation due to atmospheric gases and rain is considered to be of great influence \cite{530-17}.

The attenuation due to the absorption by oxygen and water vapour is always present, and should be included in the calculation of total propagation attenuation at frequencies above 10 GHz. The calculation method for the attenuation due to atmospheric gases is given in Recommendation ITU-R P.530-17 \cite{530-17}. Assuming that the maximum link length of the terrestrial HSR scenario we designed is 0.6 km, then the maximum value of attenuation by atmospheric gases in this case is around 0.12 dB.

Although the rain attenuation can be ignored at frequencies below 5 GHz, it must be included in attenuation calculations at higher frequencies, where its importance increases rapidly. Based on the technique for estimating long-term statistics of rain attenuation given in ITU-R P.530-17, the maximum value of the rain attenuation for terrestrial links should be no greater than 8.1074 dB.

\subsection{Excess Propagation Attenuation for Satellite-Terrestrial Links}
The excess propagation attenuation for satellite-terrestrial links is the sum of several different elements. Generally at elevation angles above $10^\circ$ (which is $45^\circ$ in this paper), only the attenuation due to atmospheric gases, rain, clouds and possible scintillation will be significant \cite{618-13}.

The gaseous attenuation which is entirely caused by absorption depends mainly on the frequency, elevation angle, altitude above sea level and water vapour density.
We can get the details for calculating the gaseous attenuation according to ITU-R P.676-11 \cite{676-11}.
Thus, the typical value of the gaseous attenuation ($A_G$) is 0.7071 dB.

Due to the uncertainty of the occurrence time and region of rainfall, it is impossible to calculate the rain attenuation accurately. After decades of observation and research, the calculation of long-term rain attenuation statistics from point rainfall rate have been summarized in ITU-R P.618-13 \cite{618-13}.
The typical value of attenuation by rain ($A_R$) is 30.0162 dB.

The attenuation due to clouds and fog has a great influence on wave propagation of satellite communication links, which can be calculated by the total columnar content of cloud liquid water in ITU-R P.840-7 \cite{840-7}.
The typical value of attenuation by clouds and fog ($A_C$) is 2.1677 dB.

The effect of the tropospheric scintillation on the signal will enhance with the increase of carrier frequency, especially above 10 GHz. A general technique for predicting the cumulative distribution of tropospheric scintillation is given in ITU-R P.618-13.
The typical value of attenuation by tropospheric scintillation ($A_S$) is 0.7638 dB.
\subsection{Total Excess Propagation Attenuation}
Since we consider two weather conditions (rainy day and sunny day) in the RT simulations, we should obtain the total excess propagation attenuation for different cases respectively.
For terrestrial links, the total excess propagation attenuation is 8.2274 dB for the rainy day and 0.12 dB for the sunny day.

For satellite-terrestrial links, the total attenuation represents the combined effect of atmospheric gases, rain, clouds and tropospheric scintillation. A general method for calculating total attenuation ($A_T$) is given by \cite{618-13}
\begin{equation}
\begin{split}
A_{T-Rainy}&=A_G+\sqrt{(A_R+A_C)^2+A^2_S}=32.90\; {\rm dB}\\
A_{T-Sunny}&=A_G+\sqrt{A^2_C+A^2_S}=3.01\; {\rm dB}
\end{split}
\end{equation}

Accordingly, the typical value of total attenuation is 32.90 dB for the rainy day and 3.01 dB for the sunny day in satellite-terrestrial links.

\section{Channel Characterization and Key Parameter Analysis}
Based on extensive RT simulation results, the channel characteristics for the terrestrial HSR system and the satellite-terrestrial system with two weather conditions are given by the following related key parameters: received power, RMS delay spread (DS), Rician $K$-factor (KF), ASA, ASD, ESA and ESD.

Based on the simulation results, it is identified that the impact of rainfall on DS, KF, ASA, ASD, ESA and ESD in this scenario setup is negligible. Thus, the impact of the weather condition on the channel characterization will mainly be considered on the SIR analysis.
If there is no special explanation, the data and results in the subsequent subsection are all obtained on the rainy weather condition.

\subsection{Received Power and Power Delay Profile}
\begin{figure}[!t]
\center
\includegraphics[width=1\columnwidth,draft=false]{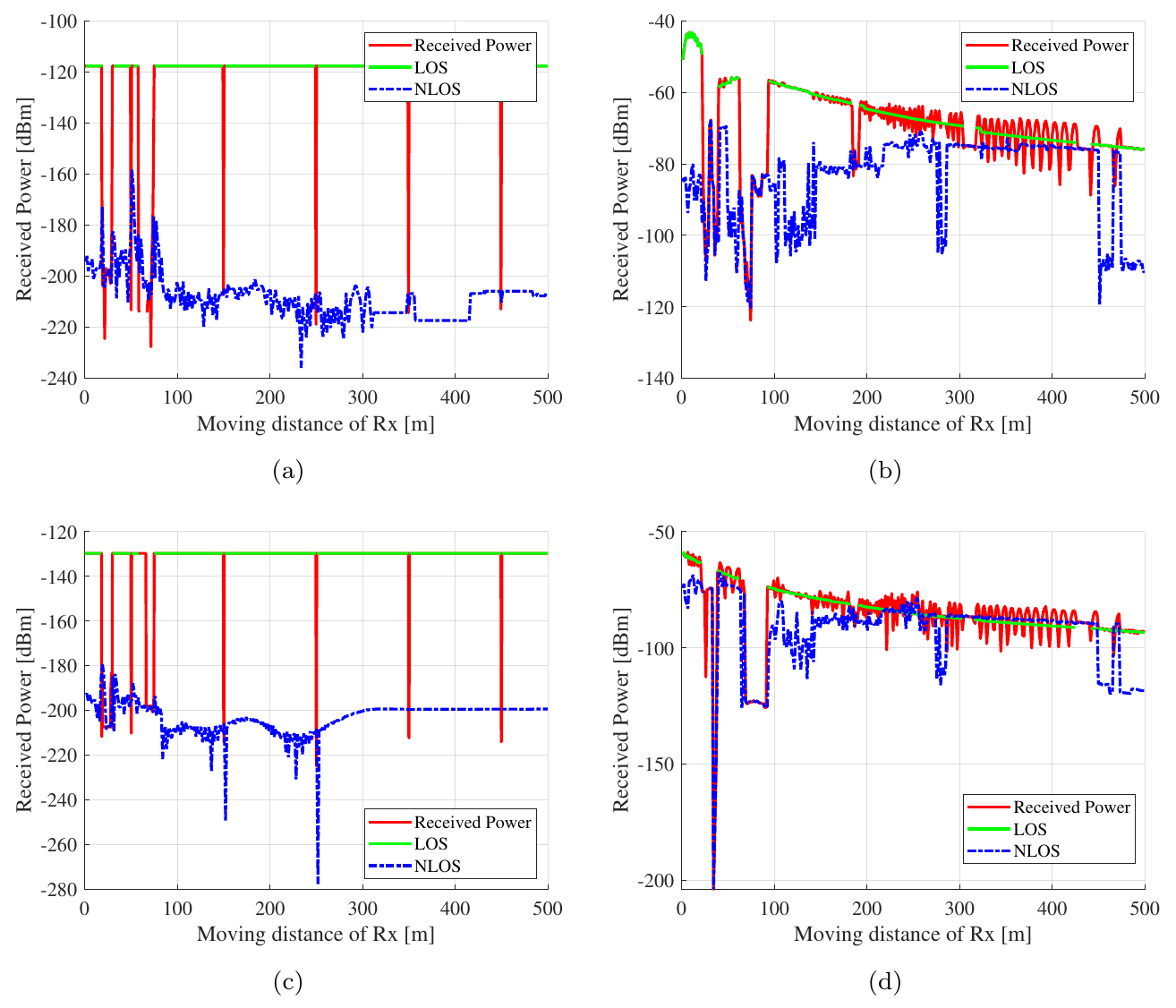}\\
\caption{Received power of four communication links: a) SA2SaUE; b) BS2TrUE; c) SA2TrUE; and d) BS2SaUE.}
\label{fig:RP}
\end{figure}

The received power for both satellite-terrestrial and terrestrial links is depicted in Fig. \ref{fig:RP}, where the green solid lines represent the direct component (i.e. LOS path), the blue dotted lines depict the ensemble of multipath components (i.e. NLOS path), and the red solid lines indicate the total received power.

For all the four links, there evidently exists deep fading in two consecutive sections where the moving distance of the Rx is approximately 20-40 m and 60-90 m, respectively. The deep fading results from the obstruction for the direct path, which is caused by crossing bridges over the railway.

\begin{figure}[!t]
\center
\includegraphics[width=0.6\columnwidth,draft=false]{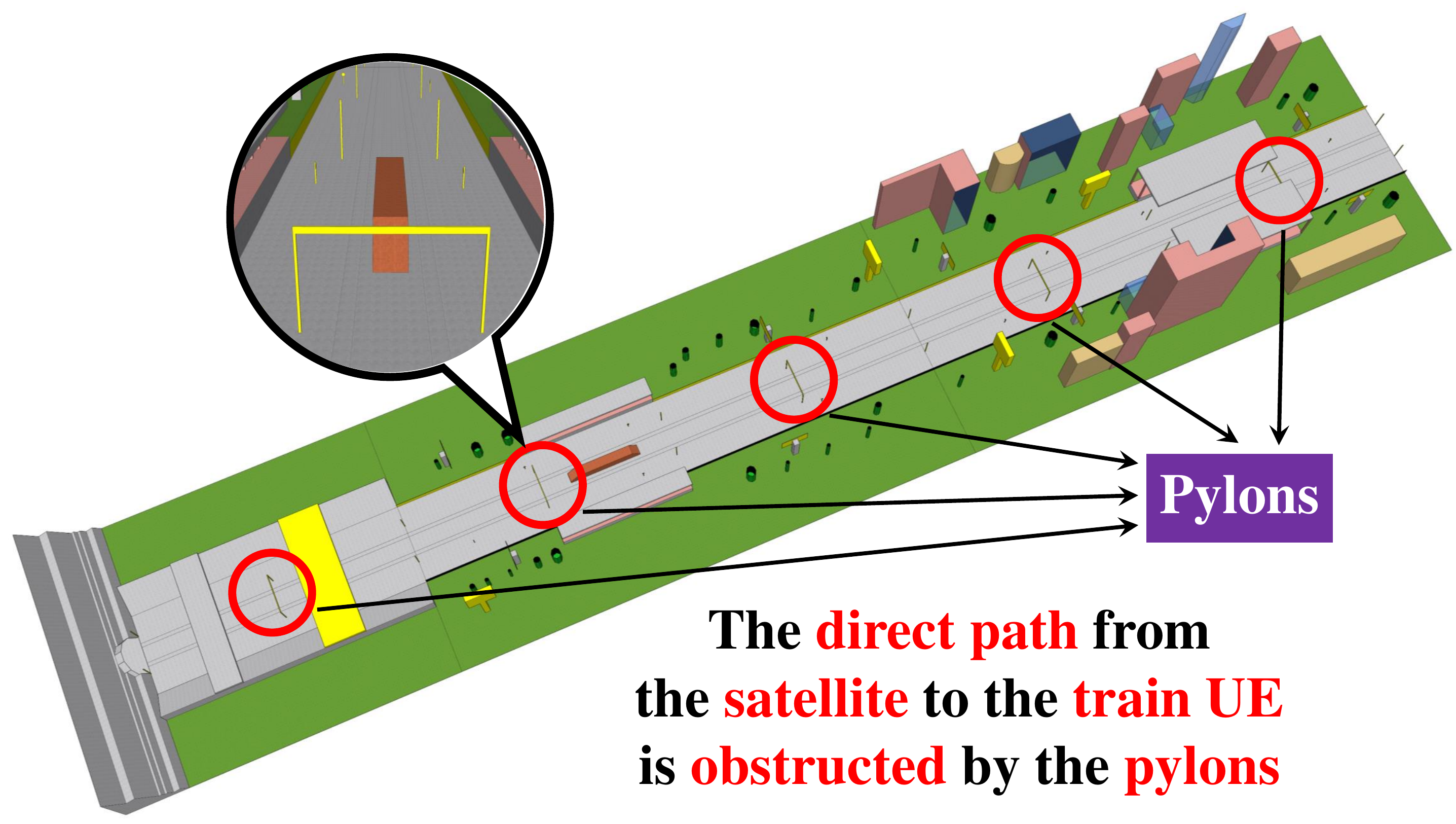}\\
\caption{Pylons in the HSR environment}
\label{fig:Pylon}
\end{figure}

As for satellite-terrestrial links (i.e. SA2SaUE and SA2TrUE), the received power is approximately a fixed value along most of the train displacement as depicted in Fig. \ref{fig:RP}(a) and Fig. \ref{fig:RP}(c).
This is because there exists a permanent direct path between the satellite and the train Rx antenna, and the impact of the multipath components on the received signal is extremely minimal due to the narrow antenna beamwidth used for satellite communications.
Moreover, compared with the TrUE, the quite narrow antenna beamwidth of the SaUE will cause more situations of the direct path obstructed by the pylons (see Fig. \ref{fig:Pylon}), and further lead to a series of deep fading at 150, 250, 350 and 450 m in Fig. \ref{fig:RP}(a) and Fig. \ref{fig:RP}(c).

\begin{figure}[!t]
\center
\includegraphics[width=0.6\columnwidth,draft=false]{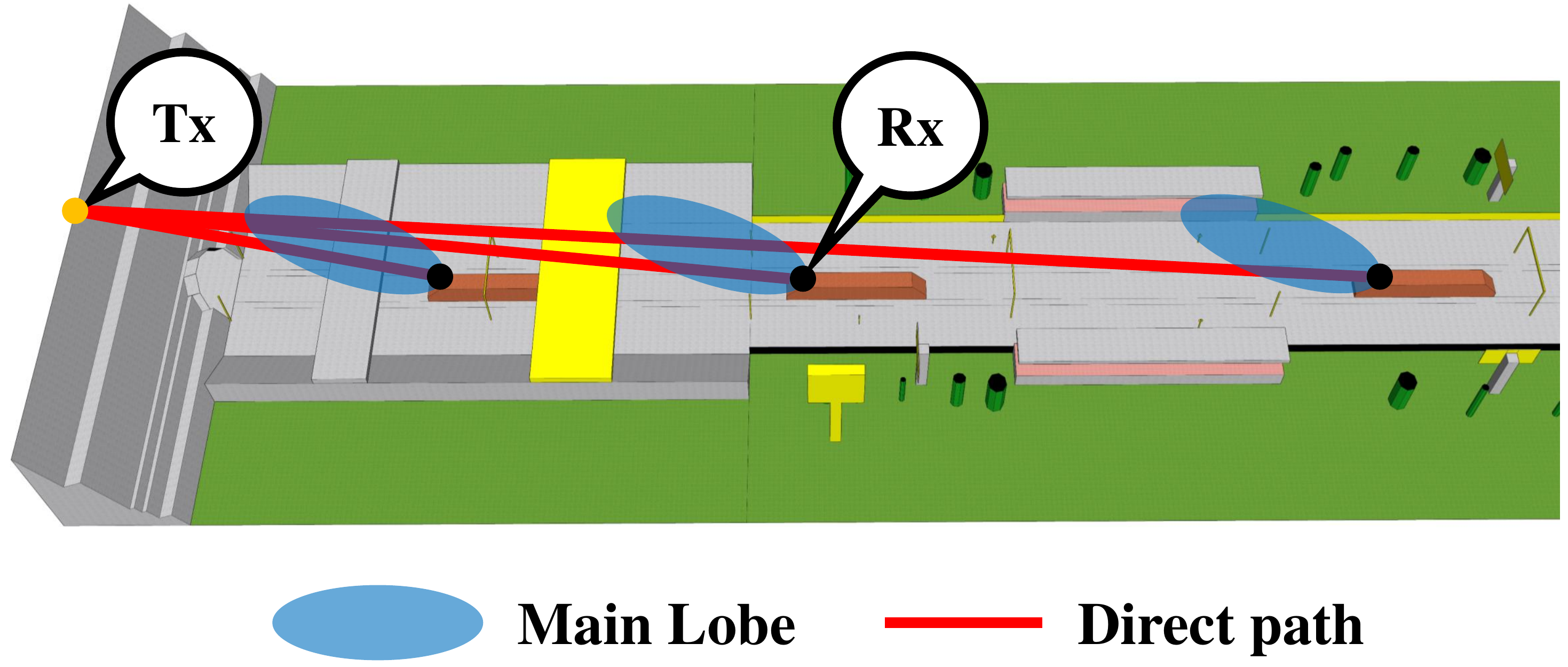}\\
\caption{The antenna misalignment as the train moves}
\label{fig:MainLobe}
\end{figure}

Furthermore, the received power of terrestrial links (i.e., BS2TrUE and BS2SaUE) decreases as the train gradually moves away from the base station.
This is not only due to the increase of the free space path loss caused by the increasing propagation distance, but also because that the direct path is not aligned with the main lobe of the Rx antenna in the elevation plane, which results in relatively low power of the direct path. This antenna misalignment is depicted in Fig. \ref{fig:MainLobe}.
Despite the fact that in terrestrial links, the LOS component is also obstructed when the train is running under the crossing bridges and pylons, it is not as pronounced as in satellite-terrestrial links, because of the reduced incidence elevation angle and the rich multipath components caused by the surrounding objects.

\begin{figure}[!t]
\center
\includegraphics[width=1\columnwidth,draft=false]{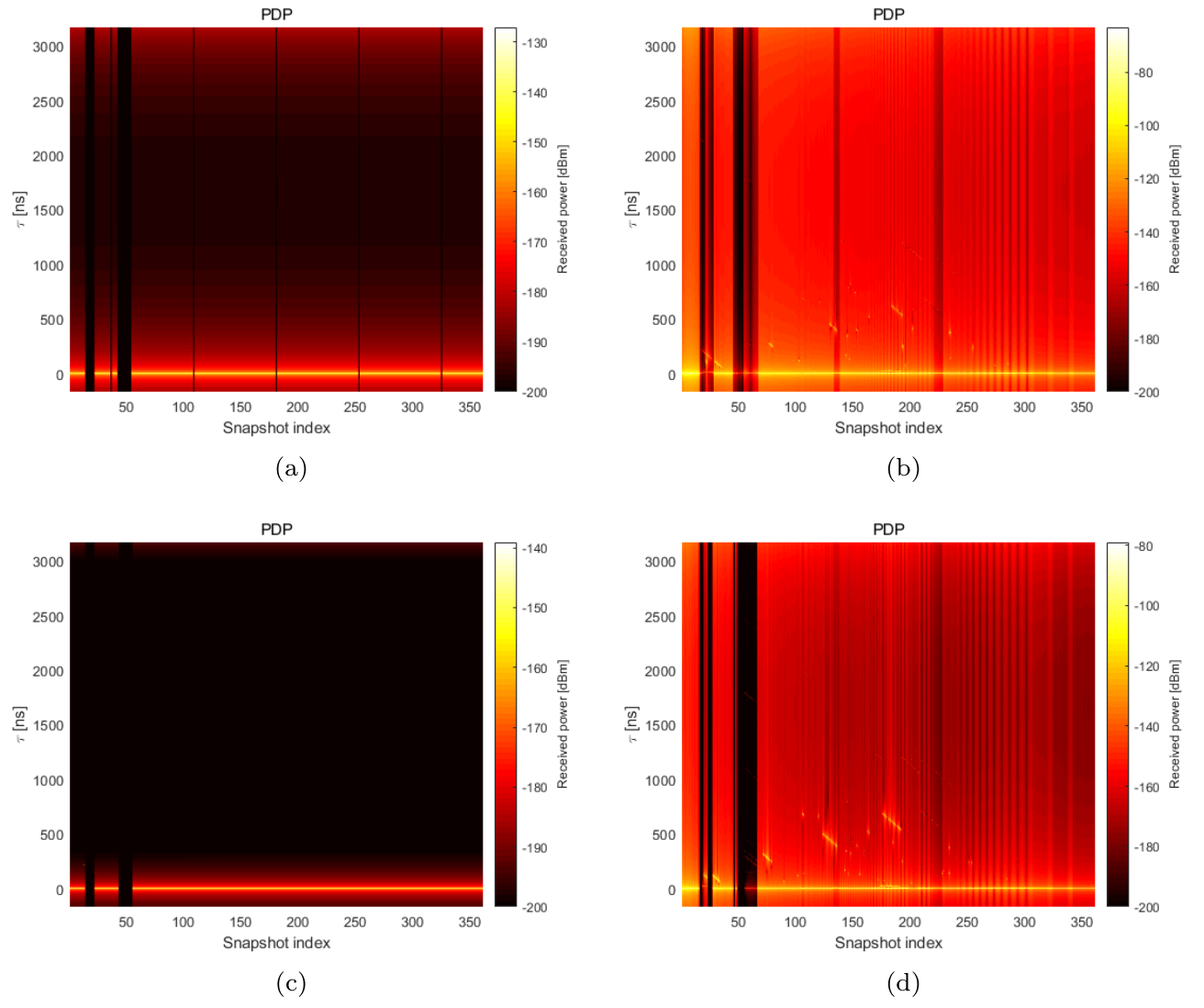}\\
\caption{PDPs of four communication links: a) SA2SaUE; b) BS2TrUE; c) SA2TrUE; and d) BS2SaUE.}
\label{fig:PDP}
\end{figure}

Moreover, the power delay profiles (PDPs) for both satellite-terrestrial links and terrestrial links are depicted in Fig. \ref{fig:PDP}.

\subsection{RMS Delay Spread}
RMS delay spread is an important measure that quantifies the dispersion effect due to propagation in the time delay domain, to which the communication systems might be sensitive. RMS delay spread is defined as the square root of the second central moment of the power delay profile (PDP) \cite{Rappaport2002Wireless} as in:
\begin{linenomath*}
\begin{equation}\label{eq:DS}
{\sigma _\tau } = \sqrt {\frac{{\sum\limits_{n = 1}^N {{\tau _n}^2 \cdot {P_n}} }}{{\sum\limits_{n = 1}^N {{P_n}} }} - {{\left( {\frac{{\sum\limits_{n = 1}^N {{\tau _n} \cdot {P_n}} }}{{\sum\limits_{n = 1}^N {{P_n}} }}} \right)}^2}}
\end{equation}
\end{linenomath*}
where ${\sigma _\tau }$ is the RMS delay spread, $P_n$ and $\tau _n$ are the power and the excess delay of the $n^{th}$ multipath, respectively.

In order to quantify the RMS delay spread for four links in the HSR environment, the obtained results are fitted by normal distribution of mean value $\mu$ and standard deviation $\sigma$. These values are depicted in Table \ref{Table:Channel_Parameters}, including the normal distribution fitting values of other key channel parameters described in the following subsections.

\begin{table*}[!t]
\centering
\caption{Extracted key channel parameters of four communication links}
\label{Table:Channel_Parameters}
\scriptsize
\begin{tabular}{c|c|c|c|c|c|c|c|c|c|c|c|c}
\specialrule{0.3pt}{2pt}{0.5pt}
\specialrule{0.3pt}{0.5pt}{2pt}
\multirow{2}{*}{\textbf{Link}} & \multicolumn{2}{c|}{\textbf{DS {[}ns{]}}} & \multicolumn{2}{c|}{\textbf{KF {[}dB{]}}}& \multicolumn{2}{c|}{$\bm{ASA\ [^{\circ}]}$} & \multicolumn{2}{c|}{$\bm{ASD\ [^{\circ}]}$} & \multicolumn{2}{c|}{$\bm{ESA\ [^{\circ}]}$} & \multicolumn{2}{c}{$\bm{ESD\ [^{\circ}]}$} \\ \cline{2-13}\rule{0pt}{8pt}
 & $\mu_{DS}$          & $\sigma_{DS}$    & $\mu_{KF}$          & $\sigma_{KF}$ & $\mu_{ASA}$              & $\sigma_{ASA}$           & $\mu_{ASD}$              & $\sigma_{ASD}$  & $\mu_{ESA}$              & $\sigma_{ESA}$           & $\mu_{ESD}$              & $\sigma_{ESD}$           \\ \hline\rule{0pt}{8pt}
BS2TrUE-R     & 0.71        & 0.48        & 26.61      & 21.96      & 0.31      & 4.06      & 0.15    & 0.18          & 3.40      & 1.77        & 0.43        & 0.22       \\ \hline\rule{0pt}{8pt}
SA2TrUE-R     & 2.41        & 0.36        & 53.44      & 7.67       & 0.02      & 0.02      & 0.16    & 0.39         & 0.06      & 0.08        & 0             & 0            \\ \hline\rule{0pt}{8pt}
BS2SaUE-R     & 3.63        & 18.72       & 26.26      & 18.98      & 3.33      & 10.84     & 0.28    & 0.28       & 3.97      & 4.88        & 0.31        & 0.15       \\ \hline\rule{0pt}{8pt}
SA2SaUE-R     & 2.42        & 0.33        & 56.76      & 1.27       & 0      & 0      & 0         & 0         & 0      & 0        & 0             & 0            \\
\specialrule{0.3pt}{1pt}{0.5pt}
\specialrule{0.3pt}{0.5pt}{2pt}
\end{tabular}
\end{table*}

\begin{figure}[!t]
\center
\includegraphics[width=1\columnwidth,draft=false]{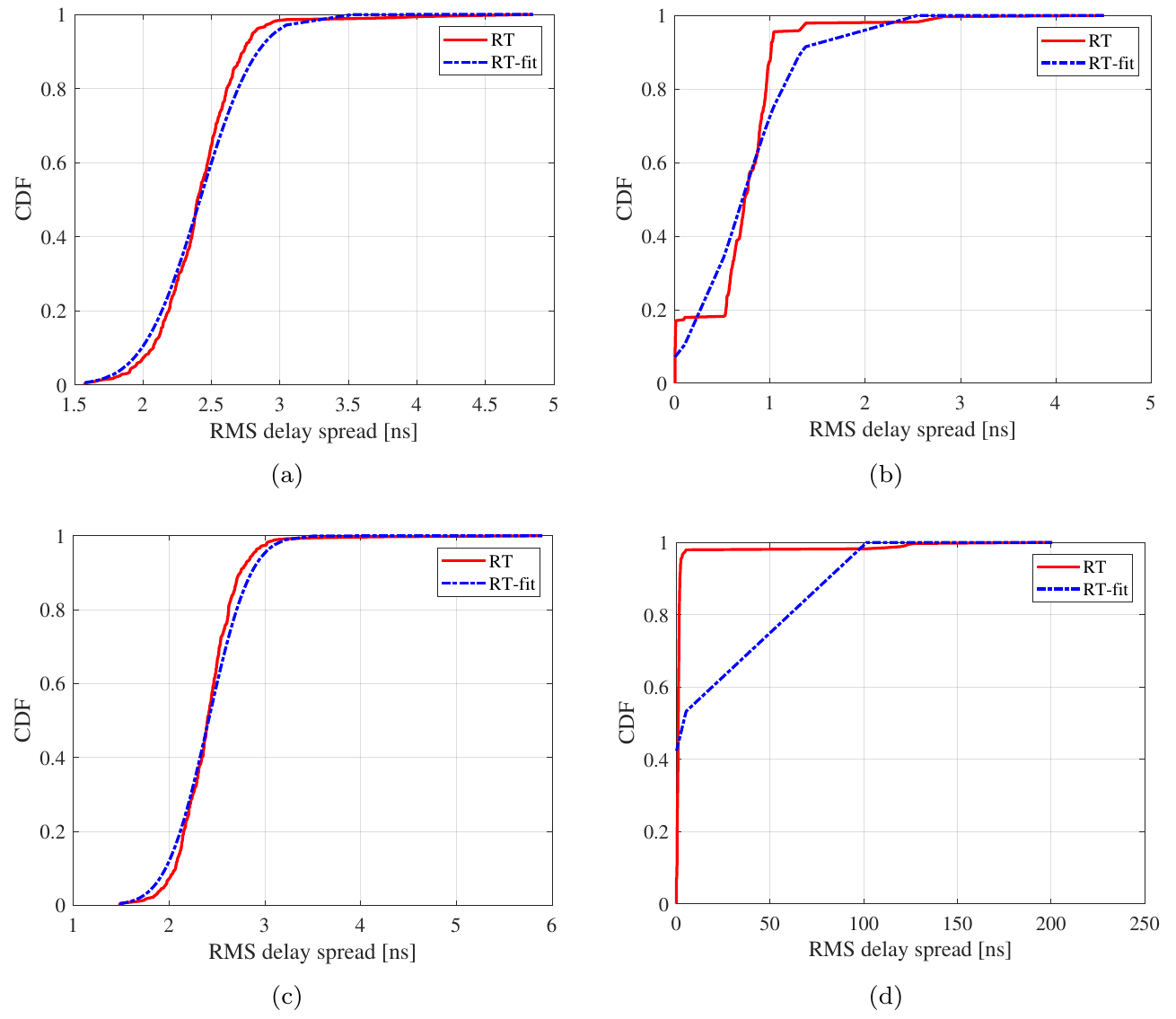}\\
\caption{RMS delay spread values of four communication links: a) SA2SaUE; b) BS2TrUE; c) SA2TrUE; and d) BS2SaUE.}
\label{fig:DS}
\end{figure}
The cumulative distribution functions (CDFs) of the RMS delay spread values for satellite-terrestrial and terrestrial communication links are depicted in Fig. \ref{fig:DS}.
As depicted in Fig. \ref{fig:DS}(a) and Fig. \ref{fig:DS}(c), the RMS delay spread of satellite-terrestrial links are similar, with values less than 3 ns with a probability of 90\%, which means that most of the strong multipath components (MPCs) are concentrated around the LOS path and the effect of MPCs on the satellite link is limited. This is consistent with the simulation results in which the scattered rays are mainly concentrated on the top of the train.

\begin{figure}[!t]
\center
\includegraphics[width=0.6\columnwidth,draft=false]{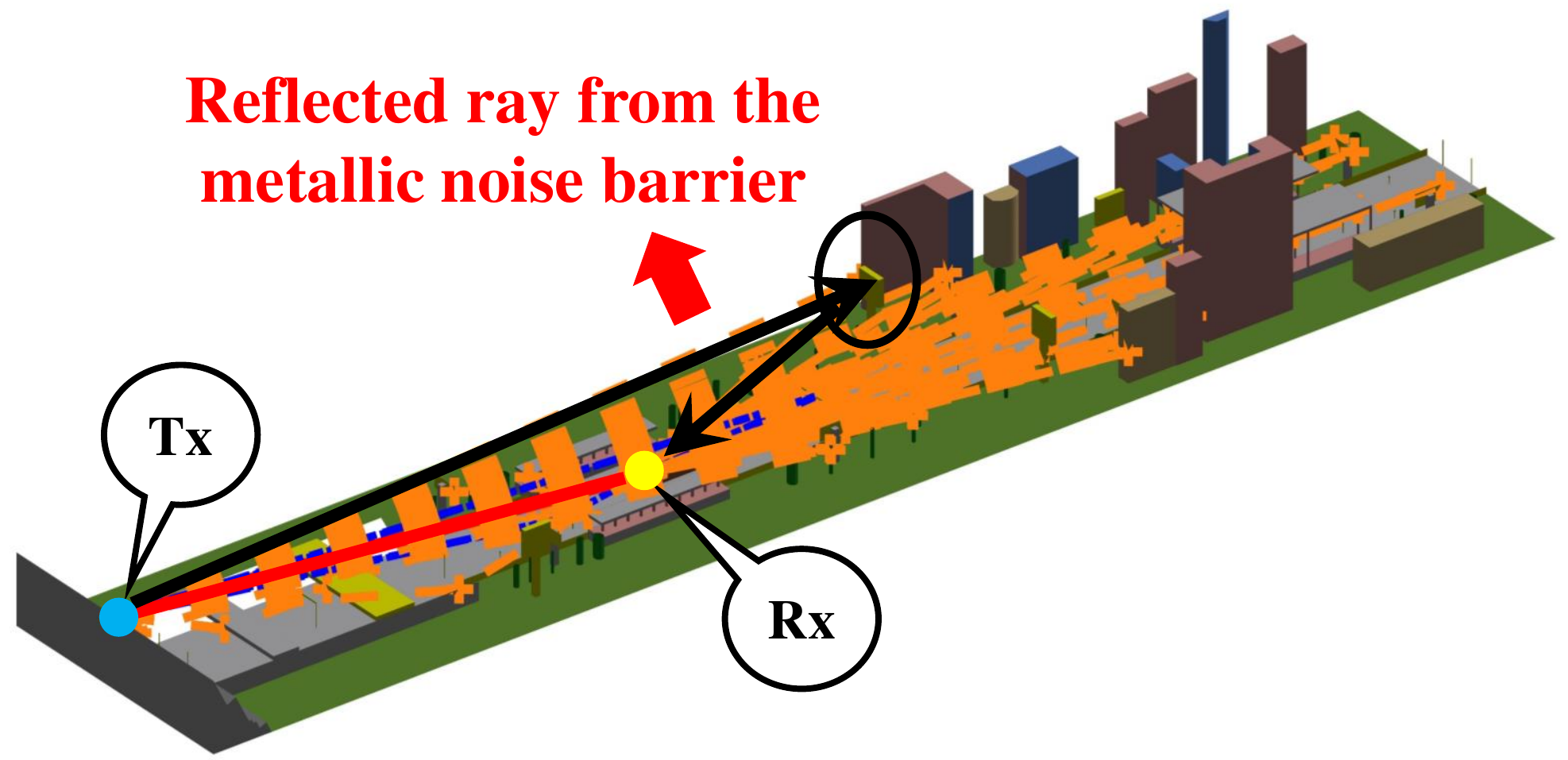}\\
\caption{One reflected path in the simulation of the BS2SaUE link}
\label{fig:RMSsnapshot}
\end{figure}

However, an unanticipated RMS delay spread value of 200 ns is found for the BS2SaUE link, as depicted in Fig. \ref{fig:DS}(d). This is because of the relatively long time delay and high signal power of a reflected path, which is caused by the distant metallic noise barrier. This reflected path can be observed in Fig. \ref{fig:RMSsnapshot}.

\subsection{Rician $K$-factor}
\begin{figure}[!t]
\center
\includegraphics[width=1\columnwidth,draft=false]{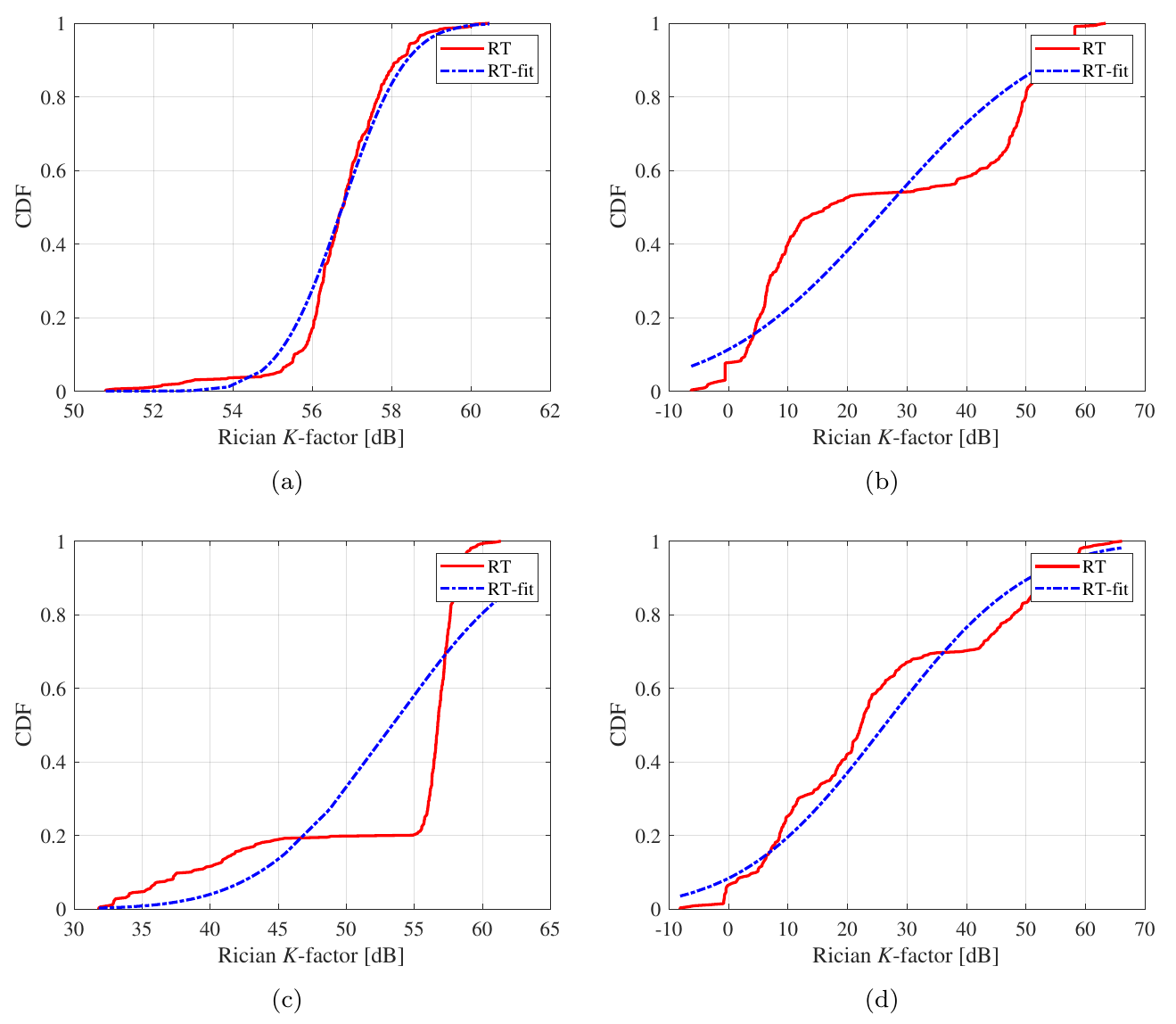}\\
\caption{Rician $K$-factor values of four communication links: a) SA2SaUE; b) BS2TrUE; c) SA2TrUE; and d) BS2SaUE.}
\label{fig:KF}
\end{figure}

The Rician $K$-factor is a significant parameter to quantify the channel fading severity, which is defined as the ratio of the power of the strongest component to the total power of the remaining components in the received signal \cite{6899647}. Thus, the Rician $K$-factor can be calculated according to its definition:
\begin{linenomath*}
\begin{equation}\label{eq:KF}
\centering
KF\left( {dB} \right) = 10 \cdot {\rm{lo}}{{\rm{g}}_{10}}    \left(\frac{{{P_{{\rm{strongest}}}}}}{{\sum {{P_{{\rm{remaining}}}}} }}  \right)
\end{equation}
\end{linenomath*}                                                                                                                                                                                                              where $KF$ is the Rician $K$-factor, ${P_{\rm{strongest}}}$ and ${P_{\rm{remaining}}}$ are the power of the strongest component and each remaining component, respectively.
The fitting results of the Rician $K$-factor are summarized in Table \ref{Table:Channel_Parameters} and the CDFs are compared in Fig. \ref{fig:KF}.

From the table and the figures, the $\mu_{KF}$ is around 55 dB for satellite-terrestrial links. The large mean values of the Rician $K$-factor are because of the strong contribution of the direct component, compared to the other rays.

However, the Rician $K$-factor values in terrestrial links are significantly smaller than that in satellite-terrestrial links, with values less than 0 dB in approximately 10\% of the recorded values. This is because of the richness of multipath components in terrestrial links, which results in an increasing ${P_{\rm{remaining}}}$.
\subsection{Angular Spread}
The four angular spreads (ASA, ASD, ESA, and ESD) are calculated through the same approach of the 3rd Generation Partnership Project (3GPP) standards \cite{3GPP}
\begin{linenomath*}
\begin{equation}\label{eq:AS}
	\sigma _{\rm{AS}} = \sqrt {\frac{{\sum\limits_{n = 1}^N {{{\left( {{\theta _{n,\mu }}} \right)}^2} \cdot {P_n}} }}{{\sum\limits_{n = 1}^N {{P_n}} }}}
\end{equation}
\end{linenomath*}
where $\sigma _{\rm{AS}}$ is the angular spread, $P_n$ is the power of the $n^{th}$ multipath component, and ${\theta _{n,\mu }}$ is defined by:
\begin{linenomath*}
\begin{equation}\label{eq:theta}
	{\theta _{n,\mu }} = \bmod \left( {{\theta _n} - {\mu _\theta} + \pi ,2\pi } \right) - \pi
\end{equation}
\end{linenomath*}
where $\theta _n$ is the AoA/AoD/EoA/EoD of the $n^{th}$ multipath and ${\mu _n}$ is calculated by
\begin{linenomath*}
\begin{equation}\label{eq:mu}
	{\mu _\theta } = \frac{{\sum\limits_{n = 1}^N {{\theta _{n }} \cdot {P_n}} }}{{\sum\limits_{n = 1}^N {{P_n}} }}
\end{equation}
\end{linenomath*}

The normal distribution fitting values of angular spreads (ASA, ASD, ESA, ESD) for each link are summarized in Table \ref{Table:Channel_Parameters}. The mean values of angular spreads are very small for satellite-terrestrial links, which indicates that the MPCs are relatively less than that in terrestrial links, and are mainly concentrated on the LOS direction.

\begin{figure}[!t]
\center
\includegraphics[width=0.65\columnwidth,draft=false]{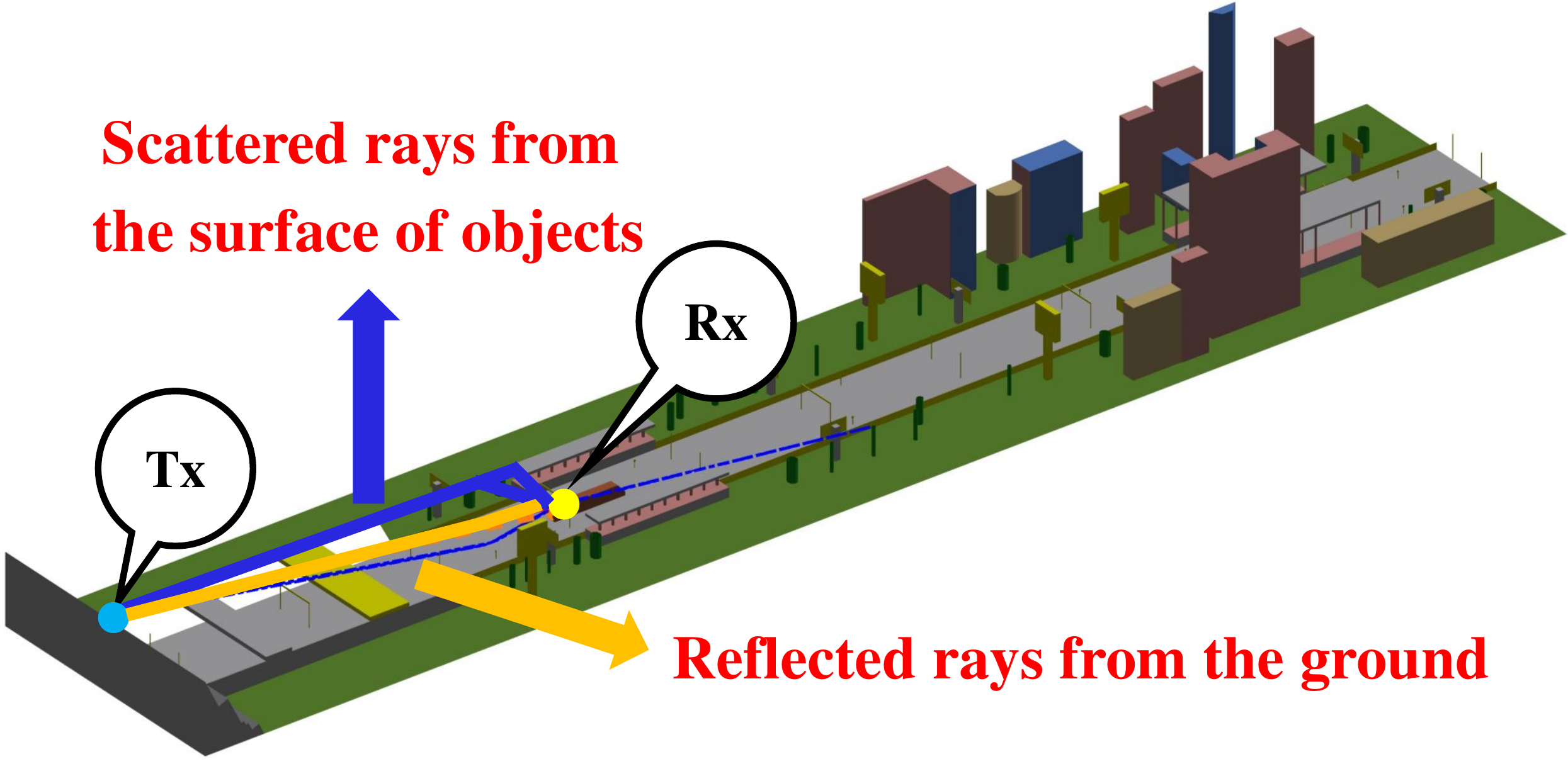}\\
\caption{Reflected and scattered rays for terrestrial links}
\label{fig:ASsnapshot}
\end{figure}

In terrestrial links, ESA and ESD values are larger than ASA and ASD, which implies that most of the multipaths components arrive from the elevation direction. This truthfully reflects that in our simulation results, a large number of reflected components are mainly from the ground, and the scattering occurs mostly on the surface of objects on both sides of the rail track. This can be observed in Fig. \ref{fig:ASsnapshot}.

\subsection{Co-channel Interference Analysis}
Frequency reuse and interference are a set of indivisible themes. Due to the fact that the satellite links and the terrestrial links use the same spectrum in this study, co-channel interference exists between these two links. Co-channel interference means the carrier frequencies of the desired signal and the interference signal are the same and are received by the receiver without discrimination, which increases the difficulty in detecting the desired signal.
Since the interference signal provides no information, this signal will contribute to a degradation of the SIR. The SIR can then be expressed as:
\begin{linenomath*}
\begin{equation}
SIR({\rm dB})=P_{\rm{signal}}({\rm dBm})-P_{\rm{interference}}({\rm dBm})
\end{equation}
\end{linenomath*}
where SIR is the signal-to-interference ratio, $P _{\rm{signal}}$ is the useful received power from the corresponding Tx, and $P_{\rm{interference}}$ is the unwanted received power from the other Tx. The SIR is then evaluated for both satellite-terrestrial and terrestrial HSR systems.

\begin{figure}[!t]
\center
\includegraphics[width=1\columnwidth,draft=false]{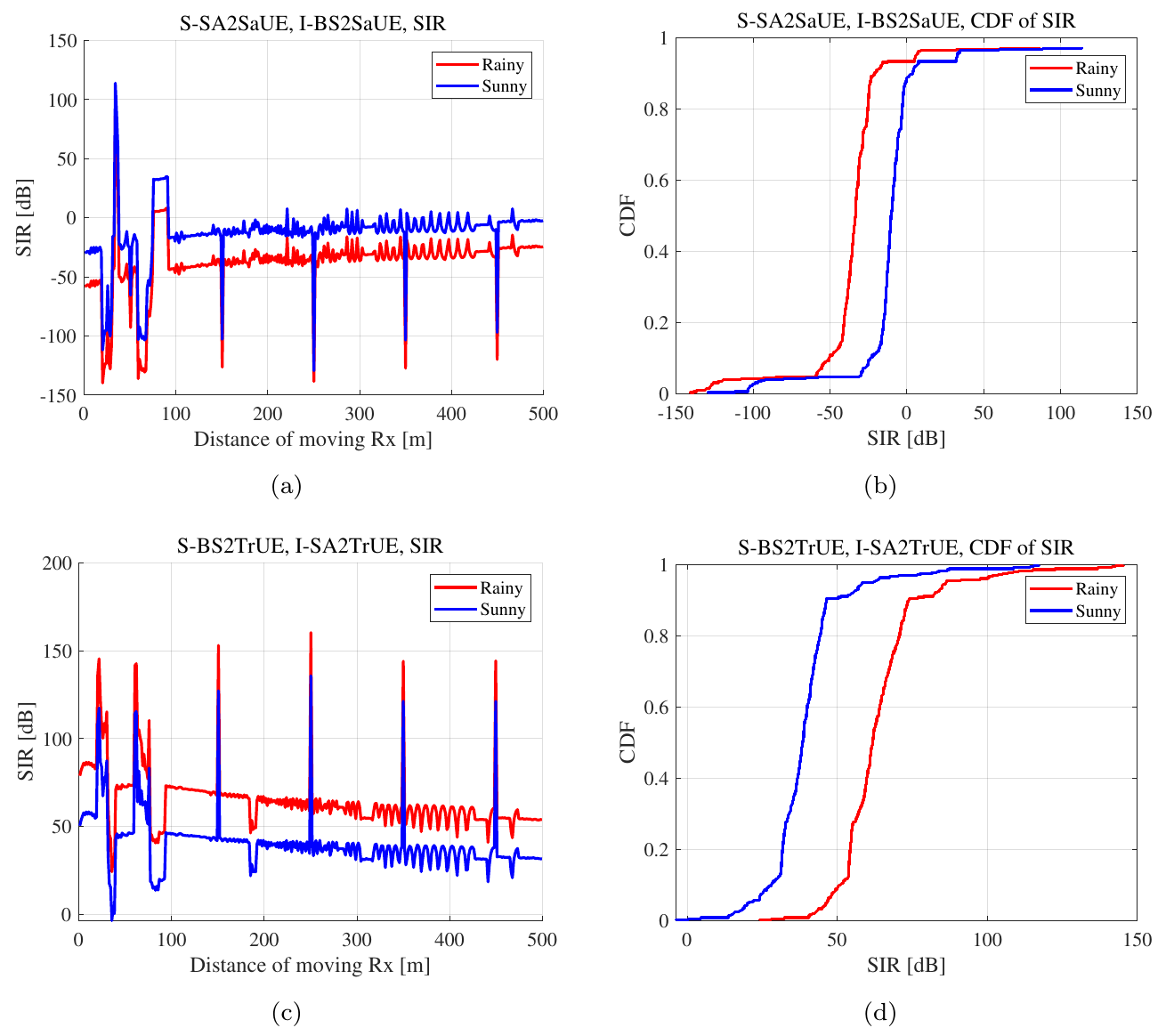}\\
\caption{Signal-to-interference ratio analysis results: a) Satellite-terrestrial system; b) CDF of satellite-terrestrial system; c) Terrestrial HSR system; and d) CDF of terrestrial HSR system.}
\label{fig:SIR}
\end{figure}

Fig. \ref{fig:SIR}(a) depicts the SIR results obtained at the satellite Rx (i.e. SaUE) in the satellite-terrestrial system. Signals from the SA2SaUE link are assumed to be useful signals while signals from the BS2SaUE link will act as interference signals.
It is obvious that the bottoms of the SIR, result from the deep fading of \emph{P$_{\rm{signal}}$}. Except for these bottoms, the SIR values are around -30 dB for rainy weather condition. Through calculating the CDF of SIR as depicted in Fig. \ref{fig:SIR}(b), the SIR coverage probability can be obtained. The probability that the SIR is higher than the threshold 0 dB is approximately 2$\%$, which indicates that the interference from terrestrial BSs will have a great impact on the effective satellite link for satellite-terrestrial communication system.

Similarly, Fig. \ref{fig:SIR}(c) presents the SIR results obtained at the terrestrial Rx (i.e. TrUE) in the terrestrial HSR system. Signals from the BS2TrUE link are assumed to be useful signals while signals from the SA2TrUE link will act as interference signals.
The peaks of the SIR occur shortly after the Rx moves below the crossing bridges. At this moment, there exists a direct path from the BS while no direct path from the satellite. On the contrary, the bottoms of SIR occur when the Rx just left the crossing bridges for there exists a direct path from the satellite while no direct path from the BS. With the exception of these peaks and bottoms, the SIR values are around 60 dB for rainy weather condition, which are marked in Fig. \ref{fig:SIR}(c). Through calculating the CDF of SIR as depicted in Fig. \ref{fig:SIR}(d), the probability that the received SIR is greater than 40 dB is approximately 98$\%$, which is reliable enough for future intelligent rail transportation applications. Evidently, the interference from satellite antennas will not have much impact on the effective terrestrial link for terrestrial HSR system.

\subsection{Effect of Weather Conditions on SIR}
According to the calculation methods for estimating long-term rain attenuation statistics on terrestrial and satellite-terrestrial communication links given in ITU-R P.530-17 and ITU-R P.618-13, the attenuation due to rain is proportional to the distance of the communication link. Accordingly, the rainfall has a greater impact on the received signal from satellite antennas, compared with the received signal from terrestrial BSs.

Thus for the terrestrial HSR system, the attenuation of the interference signal caused by rainfall is greater than that of the useful signal, which leads to higher SIR values in the rainy day. While as for the satellite-terrestrial system, the attenuation of the useful signal caused by rainfall is greater than that of the interference signal, which results in lower SIR values in the rainy day. This is depicted in Fig. \ref{fig:SIR}.

\section{Conclusion}
In this paper, the satellite-terrestrial channel at 22.6 GHz is characterized for a typical HSR environment. The CloudRT platform is used to extract the key channel parameters of a realistic 3D HSR scenario. The objects in the scenario are defined and reconstructed according to their typical geometries and materials. Channel characterization and respective conclusions are drawn based on simulation results.

Obstacles that commonly appear in HSR scenarios, such as crossing bridges and pylons, will severely affect the performance of the wireless communication systems, since they can block the LOS path and therefore, cause deep fading on the received power.
Compared with terrestrial links, this phenomenon becomes more pronounced in satellite-terrestrial links, since for satellite links, the LOS path contributes more than other multipath components, which in turns makes satellite links more sensitive for shadowing.

The maximum value of the rain attenuation for terrestrial links of the scenario in this paper should be no greater than 8 dB, and the typical value of the rain attenuation for satellite-terrestrial links is around 30 dB.
Through analyzing the channel parameters for different weather conditions, we conclude that the rainfall would not influence channel parameters like Rician $K$-factor, RMS delay spread, and angular spreads, but it will influence the received power and corresponding interference between terrestrial link and satellite-terrestrial link.

When the Rx antenna is mounted on the top of the train, the large-scale objects like buildings and train stations will provide strong reflected and scattered contributions which have significant influence on the wireless channel, while some small-scale objects such as billboards and traffic signs should also not be neglected since they have a great impact on channel parameters such as Rician $K$-factor and RMS delay spread.

The SIR between satellite-terrestrial and terrestrial communication systems is also analyzed.
The SIR values for satellite-to-terrestrial interference are around 60 dB, which indicates that the interference from satellite antennas will not have much impact on the effective terrestrial links. This basically meets the requirement of good communication performance in 5G mmWave channel. On the contrary, the SIR values for terrestrial-to-satellite interference are around -30 dB, which states that satellite-terrestrial communication links will be severely affected by the existence of terrestrial links, since the strength of the interference from terrestrial BSs is comparable to the received useful signal from satellite antennas.

The channel characterization analysis and the key channel parameter extraction provided in this paper, are suitable for effective link budget of the satellite-terrestrial channel in a realistic HSR environment, which will help the research community understand the propagation channel when designing mmWave technologies and communication system for future intelligent rail transportation.
Future work will address the mmWave satellite-terrestrial channel characterization in more scenarios and potential system configurations.

\acknowledgments
This work was supported in part by Institute of Information \& communications Technology Planning \& Evaluation (IITP) grant funded by the Korea government(MSIT) (No.2018-0-00792, QoE improvement of open Wi-Fi on public transportation for the reduction of communication expense), and in part by IITP grant funded by the Korea government(MSIT) (No.2018-0-00175, 5G AgiLe and fLexible integration of SaTellite And cellulaR). Readers can access the data of this paper in the 4TU.Centre for Research Data by the link {\color{blue} \underline{\emph{https://doi.org/10.4121/uuid:f6c34e2e-fa34-4bd6-9046-1c350a9bb5db}}}.


%
%

\bibliography{reference}

%
%
%
%
%

\end{document}